\documentclass[fleqn,usenatbib]{mnras}

\usepackage{newtxtext,newtxmath}

\usepackage[T1]{fontenc}

\DeclareRobustCommand{\VAN}[3]{#2}
\let\VANthebibliography\thebibliography
\def\thebibliography{\DeclareRobustCommand{\VAN}[3]{##3}\VANthebibliography}

\usepackage{graphicx}	
\usepackage{amsmath}	
\usepackage{longtable}
\usepackage{threeparttablex}
\usepackage{ulem}
\usepackage{xcolor}
\newcommand{\blackhole}{4U~1543$-$47}

\title[Peculiar radio-bright behaviour in 4U~1543$-$47]{Peculiar radio-bright behaviour of the Galactic black hole transient 4U~1543$-$47 in the 2021--2023 outburst}

\author[X. Zhang et al.]{
X. Zhang,$^{1,2}$
W. Yu,$^{1}$\thanks{E-mail: wenfei@shao.ac.cn}
F. Carotenuto,$^{3}$
S.E. Motta,$^{4}$
R. Fender,$^{3,5}$
J. C. A. Miller-Jones,$^{6}$
\newauthor
T. D. Russell,$^{7}$
A. Bahramian,$^{6}$
P. Woudt,$^{5}$
A. K. Hughes$^{8}$
and G. R. Sivakoff$^{8}$
\\
$^{1}$Shanghai Astronomical Observatory, Chinese Academy of Sciences, 80 Nandan Road, Shanghai 200030, China\\
$^{2}$University of Chinese Academy of Sciences, 19A Yuquanlu, Beijing 100049, China\\
$^{3}$Department of Physics, University of Oxford, Denys Wilkinson Building, Keble Road, Oxford OX1 3RH, UK\\
$^{4}$Istituto Nazionale di Astrofisica, Osservatorio Astronomico di Brera, via E.\,Bianchi 46, 23807 Merate (LC), Italy\\
$^{5}$Department of Astronomy, University of Cape Town, Private Bag X3, Rondebosch 7701, South Africa\\
$^{6}$International Centre for Radio Astronomy Research - Curtin University, GPO Box U1987, Perth, WA 6845, Australia\\
$^{7}$INAF, Istituto di Astrofisica Spaziale e Fisica Cosmica, Via U. La Malfa 153, I-90146 Palermo, Italy\\
$^{8}$Department of Physics, University of Alberta, CCIS 4-181, Edmonton, AB T6G 2E1, Canada\\
}

\date{Accepted XXX. Received YYY; in original form ZZZ}

\pubyear{2022}

\begin{document}
\label{firstpage}
\pagerange{\pageref{firstpage}--\pageref{lastpage}}
\maketitle

\begin{abstract}

Correlated behaviours between the radio emission and the X-ray emission in Galactic black hole X-ray binaries (BH XRBs) in the X-ray hard state are crucial to the understanding of disc-jet coupling of accreting black holes. The BH transient \blackhole~went into outburst in 2021 following $\sim$~19 years of quiescence. We followed it up with $\sim$~weekly cadence with MeerKAT for about one year and a half until it faded into quiescence. Multi-epoch quasi-simultaneous MeerKAT and X-ray observations allowed us to trace the compact jet emission and its X-ray emission. In its hard spectral state across three orders of magnitude of X-ray luminosities above $\sim$~10$^{34}$ ergs\,{s}$^{-1}$, we found the correlation between radio and X-ray emission had a power-law index of 0.82$\pm$0.09, steeper than the canonical value of $\sim$~0.6 for BH XRBs. In addition, the radio vs.\ X-ray correlation show a large range of the power-law normalization, with the maximum significantly larger than that obtained for most BH XRBs, indicating it can be particularly radio-bright and variable in the X-ray binary sample. The radio emission is unlikely diluted by discrete jet components. The observed peculiar radio-bright and variable behaviours provide the evidence for the relativistic effects of a variable Lorentz factor in the range between 1 and $\sim$~2 of the compact jet.

 \end{abstract}

\begin{keywords}
radio continuum: transients -- X-rays: binaries -- radio: jets
\end{keywords}



\section{Introduction}
Galactic Black Hole X-ray Binaries (BH XRBs) are stellar binary systems, wherein a central compact object, a stellar mass black hole (typically $\lesssim$~20 $M_{\odot}$), accretes matter from a companion star and forms an accretion disc around the former. Galactic BH XRBs are transients in most cases, spending large portions of their life time in quiescence (typically $\lesssim$~10$^{34}$ ergs\,{s}$^{-1}$) and occasionally interrupted by periods of outbursts \citep[e.g.,][]{1997ApJ_Chen,Yan_2015}, usually found by all sky X-ray monitors and/or other sensitive X-ray telescopes. Galactic BH XRBs typically show specific timing and spectral properties \citep[e.g.,][]{Remillard_2006,Belloni_2016} for each of the spectral states, i.e., hard state, hard and soft intermediate states, soft state \citep[e.g.,][]{2001ApJS_Homan,Belloni_2010}. These properties constitutes the general picture of coupling between the generation of jets and the accreting processes (i.e., disc-jet coupling): During the initial rising in X-ray flux of an outburst, they reside in a so-called hard spectral state where the X-ray emission is attributed to the comptonization of soft X-ray photons from a radiatively inefficient accretion flow by a corona, described by a power-law X-ray spectrum. At this stage, there exists a continuously launched compact jet, which is thought to be originated from the superposition of multiple self-absorbed synchrotron components \citep[e.g.,][]{1979ApJ_Blandford,2000ApJ_Dhawan,Fender_2001,2001MNRAS_Stirling}, presenting an optically thick flat or slightly inverted radio spectrum ($\alpha \gtrsim 0$ in the form of $S_{\nu} \propto \nu^{\alpha}$, where $S_{\nu}$ is the observed flux density at the frequency $\nu$; e.g.\ \citealt{2000MNRAS_Fender,Fender_2001}) in the radio regime and extending to the infrared. As the X-ray flux gets higher and higher, reaching a point that is even close to the Eddington luminosity, and the X-ray spectrum progressively softens, they will be going through hard and soft intermediate states when discrete ejection(s) with optically thin steep radio spectra ($\alpha$~$\sim$~$-$0.7) would be launched \citep[e.g.,][]{1994_Mirabel_Nature,1999_Fender,2012MNRAS_Miller-Jones,Russell_2019,Bright_2020}, which can be followed up at a late time by sensitive radio telescopes. These ejections had been associated with most powerful radio flaring events \citep{1995_Hjellming_Nature,1999_Fender,Bright_2020}. The systems then move towards to the soft state, wherein the X-ray emission is thermal, described by a geometrically thin, optically thick accretion disc, while the compact jet is strongly suppressed (e.g.,\ \citealt{Bright_2020}). As the mass accretion rate drops and the disc cools after Galactic BH XRBs reaching their peak luminosities, the outburst begins to fade. The systems then transition back to the hard state through the intermediate states, features replenished compact jet. Lastly, the sources go back to the quiescent state. A typical outburst is usually expected to last from weeks to months, and in some cases even years \citep[e.g.,][]{1997ApJ_Chen,Yan_2015}. Therefore BH XRBs are ideal to study the disc-jet coupling and the production of jets, because they show spectral and timing variability on a wide range of accessible time-scales from sub-seconds to years, 
while other black hole systems such as AGNs, evolve on much longer time-scales.

During the hard and quiescent states of Galactic BH XRBs, the radio luminosity, 
produced by the compact jets, is non-linearly correlated to the X-ray luminosity, which primarily probes the inner accretion flow of the central engines \citep{1998A&A_Hannikainen,2000A&A_Corbel,2003A&A_Corbel,Gallo_2003,Gallo_2014}, in the form of $L_R\propto L_X^\beta$ ($L_R$ and $L_X$ are measured luminosities in radio and X-ray). Such a correlation has become an unparalleled observational evidence supporting and understanding the disc-jet coupling in BH XRBs (\citealt{Fender_2004}) and has been found to extend to black holes of all mass scales (e.g., \citealt{2003MNRAS_Merloni}), forming a fundamental plane. While in Neutron Star X-ray Binaries (NS XRBs) the correlation seems to scatter from source to source, Galactic BH XRBs show distinct two branches, the standard branch ($L_R\propto L_X^{\sim 0.6}$; \citealt{2013MNRAS_Corbel}) and the `outlier branch' (or `radio-quiet' branch with $L_R\propto L_X^{\sim 1.4}$; e.g.,\ \citealt{2004ApJ_Corbel,Rodriguez_2007,2012MNRAS.423..590G}), though a comprehensive study of the BHs can be broadly described as a single correlation \citep{2018MNRAS.478L.132G}. Notably, some BHs have been found to switch from a high X-ray luminosity `outlier' track to a low X-ray luminosity standard track \citep[e.g.,][]{2010MNRAS.401.1255J,2011MNRAS.414..677C,2021MNRAS.504..444C,2021MNRAS.505L..58C,2021MNRAS.501.5776M}. On the one hand, observations have shown that BH spin was not found to correlate with the production of jets or play a role in causing the observed standard branch:`outlier branch' dichotomy \citep{2010MNRAS_Fender,2018MNRAS_Espinasse}. Similarly, system parameters like inclination angle \citep{2018MNRAS_Espinasse,Motta_2018,2011MNRAS_Paolo} of jets, orbital period, disc size \citep{2011MNRAS_Paolo} were not found to robustly determine the two branches either. However, it's found that they are statistically belong to two groups of radio spectral index (with mean value of $\alpha=$~0.2 and -0.2 for standard and `outlier branch'; \citealt{2018MNRAS_Espinasse,2012MNRAS.423..590G}), which may give insights into the underlying mechanism of the two branches and production of jets (although still uncertain). Notably, \cite{2015MNRAS.450.1745R} suggests that the variation of Lorentz factor of the compact jets of MAXI J1836$-$194 caused the observed `outlier branch' peculiarity. One the other hand, there are some physical models proposed to explain the observed dichotomy, for example, variations in the jet magnetic field \citep{Casella_2009}, difference in accretion flow radiative efficiency or the dependence of accretion energy injected into jets on accretion rate \citep{Coriat_2011}, comptonization of additional photons from a weak, cool inner disc \citep{Meyer_Hofmeister_2014}.

\subsection{4U 1543--47 and its 2021--2023 outburst}
\label{sec:1_1} 
\blackhole~(IL Lup), a BH XRB system that hosts a BH with mass of 9.4$\pm$1.0 M$_{\odot}$ and a companion star with mass of 2.45$\pm$0.15 M$_{\odot}$ \citep{2003A&A_Ritter}, was firstly discovered in an X-ray outburst back in 1971 \citep{1972ApJ} by {\it Uhuru} satellite. Since then, the source had been detected in outbursts in 1983 \citep{1984PASJ}, 1992 \citep{1992IAUC} and 2002 \citep{2002IAUC}. The distance to the BH XRB system is determined to be 7.5$\pm$0.5~kpc \citep{2002AAS_Orosz} through modeling the optical light curves and spectra of the system. Independently, a distance of 5$^{+2.0}_{-1.2}$ kpc is derived based on the measured optical parallax of the source (listed in Gaia Data Release 3) and the assumption of the prior of low mass BH XRBs~\citep{2019MNRAS.489.3116A}. The system has an inclination angle of 20.7$^{\circ}\pm$1.5$^{\circ}$ \citep{2003IAUS_Orosz}, while \citet{2020MNRAS_Dong} found an inclination angle of the inner disc as $\sim36.3^{\circ}$. 

On 2021 June 11 the source was found in a new X-ray outburst by the Monitor of All-sky X-ray Image/Gas Slit Camera (hereafter \textit{MAXI}; \citealt{2009PASJ_Matsuoka}). We started to monitor this source in radio 8 days after the discovery of the outburst with Meer Karoo Array Telescope (MeerKAT; \citealt{2009IEEEP_Jonas}) with $\sim$~weekly cadence, as part of the ThunderKAT\footnote{The HUNt for Dynamic and Explosive Radio transients with MeerKAT: \url{http://www.thunderkat.uct.ac.za}} Large Survey Programme \citep{fender2017thunderkat}, which observes and monitors relativistic jet activities from many BH and NS transients, some of them are detected for the first time in radio \citep[e.g.,][]{Russell_2019,Bright_2020,2021MNRAS.504..444C,2021MNRAS.501.5776M,2022MNRAS.510.1258Z,2023ApJ...948L...7B}. In addition, Australia Telescope Compact Array (ATCA) observations were included during the hard state evolution of the outburst. We also made use of X-ray observation data sets: X-ray Telescope (\textit{XRT}) instrument on board the Neil Gehrels Swift Observatory (\textit{Swift}; \citealt{2004ApJ_Gehrels}), as part of SWIFTKAT project in ThunderKAT collaboration, as well as the archival \textit{XRT} observations available during the source's 2021--2023 outburst; daily monitoring X-ray data from the Burst Alert Telescope (\textit{BAT}) on board the Neil Gehrels Swift Observatory and \textit{MAXI}; pointed observation data taken from the Neutron Star Interior Composition Explorer (\textit{NICER}, \citealt{2016SPIE_Gendreau}) on board the International Space Station. In this paper, we report our study of the correlated behaviour of the radio core jet emission and the X-ray emission from the accretion flow in \blackhole. The investigation of the episodic jets in \blackhole\ is reported in a separated paper elsewhere (Zhang et al. 2024, to submit). 

\begin{figure}
	\includegraphics[width=\columnwidth]{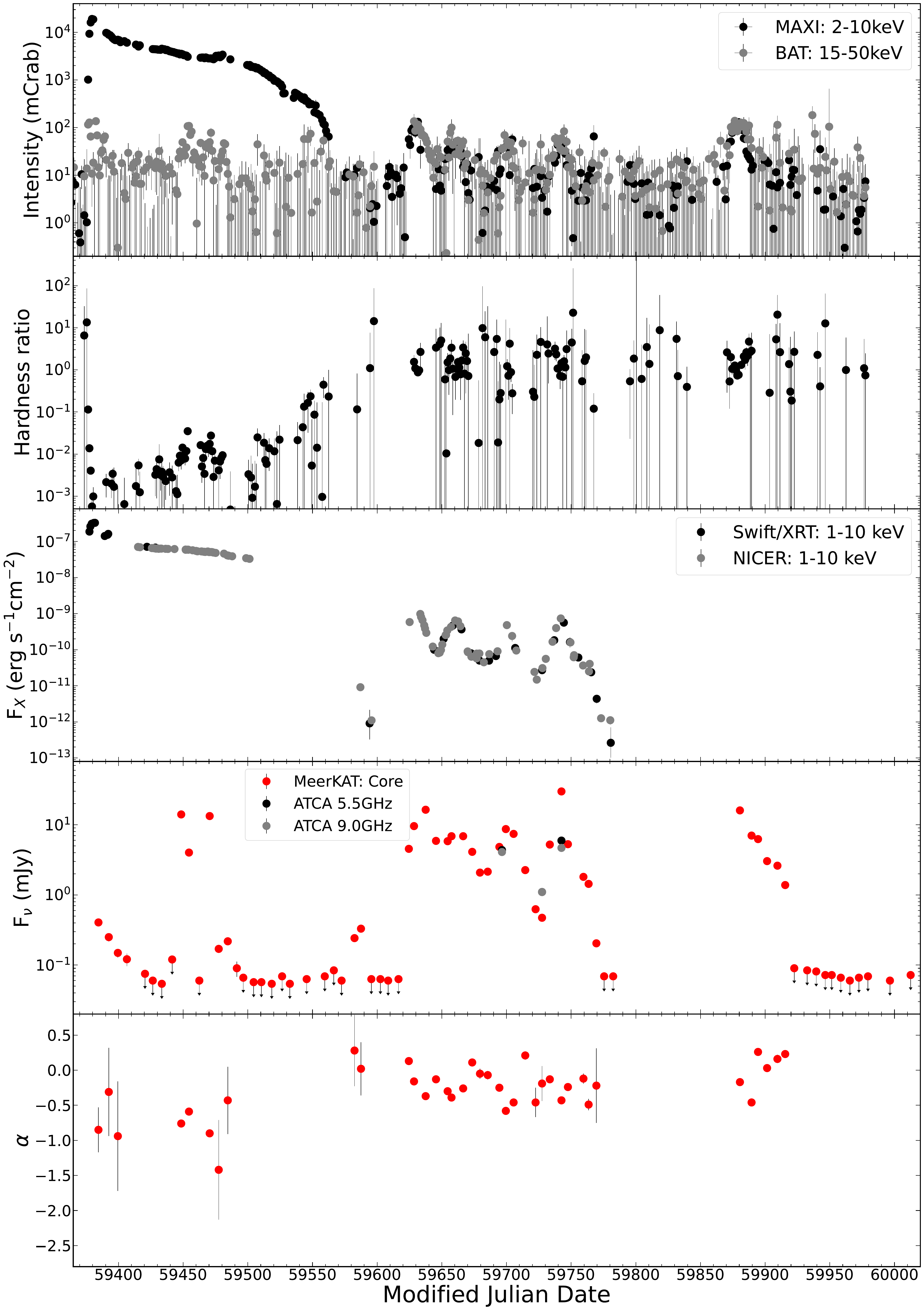}
    \caption{The radio and X-ray multi-instrument light curves. \textit{First panel}: \textit{MAXI} 2--10 keV one-day-averaged light curve (shown with black filled circles) and \textit{BAT} 15--50 keV one-day-averaged light curve (shown with gray filled circles) of \blackhole~throughout the 2021--2023 outburst. \textit{Second panel}: Hardness ratio calculated as \textit{BAT} 15--50 keV intensity over \textit{MAXI} 2--10 keV intensity. \textit{Third panel}: \textit{NICER} (gray filled circles) and \textit{Swift/XRT} (black filled circles) light curves of~\blackhole. \textit{Fourth panel}: MeerKAT radio detections (red filled circles) and 3 $\sigma$ non-detections (red filled circles with arrows) of the core radio jet emission, radio detections with ATCA at 5.5 GHz (black filled circles) and 9 GHz (gray filled circles). \textit{Fifth panel}: Radio spectrum of the core radio jet emission when detections at two sub-bands (1.07 GHz and 1.5 GHz) were obtained.}
    \label{fig:MAXI_BAT_Swift_Nicer_MeerKAT_4U1543}
\end{figure}

\section{Observations}
\subsection{Radio observations}
\subsubsection{MeerKAT observations}
\label{sec:2_1}
MeerKAT is a precursor array telescope of the Square Kilometre Array (SKA). 
The telescope in the current phase consists of 64 antennas each with diameter of 13.5 meters. 
We started to monitor the BH 4U 1543$-$47 under the ThunderKAT XRB project with $\sim$ weekly cadence with MeerKAT since 2021 June 19, 8 days after the starting of the source's X-ray outburst, until 2023 March 9. During the monitoring observations, each of them was taken at a central frequency of 1.28 GHz (\textit{L}-band) with 4096 channels in total reaching a bandwidth of 856 MHz, while the integration time is 8 seconds. We observed J1939$-$6342 for 5 min at the beginning of each run to calibrate the absolute flux and bandpass. J1501$-$3918 was used as the phase calibrator and was observed for 2 min before and after a 15-min on-source exposure time on \blackhole. Data reduction 
(i.e., flagging, calibration) was completely done through Common Astronomy Software Application package (\textsc{casa} 5.6.2-3, hereafter \textsc{casa}; \citealt{2007ASPC_McMullin,2022PASP..134k4501C}). Then we did a further flagging for the calibrated target data using \textsc{rflag} in \textsc{casa} to edit bad data that still remained. Finally, after phase-only self calibration, we imaged the data in total intensity (Stokes I) using \textsc{tclean} in \textsc{casa}. We used a Briggs weighting scheme \citep{1995PhDT.......238B} with a robust parameter of 0 to achieve compromise between the side-lobe effects across the field and sensitivity. 
We made use of the \textsc{casa} task \textsc{imfit} to measure the flux density of the source for all epochs. The local rms noise in each observation was obtained from a close-by source-free region. The source detections and corresponding flux densities of the source were also cross-checked by both the dedicated \textsc{oxkat} semi-automated pipeline \citep{2020ascl.soft_oxkat} and the Science Data Processor (SDP) pipeline images, all of which show consistency with our measurements as introduced above. Details, including the measured flux densities and rms noises are shown in supplementary information, 
also see Figure~\ref{fig:MAXI_BAT_Swift_Nicer_MeerKAT_4U1543} for the radio light curve.

\subsubsection{ATCA observations}
The Australia Telescope Compact Array (ATCA) observed \blackhole~on three dates during the 2021--2023 outburst (under project code CX501, PI: Russell). Observations were taken on 2022 April 27 17:37:00--22:03:40 UT (MJD~59696.83$\pm$0.09), 2022 May 28 11:24:50--13:33:30 UT (MJD~59727.52$\pm$0.04), and 2022 June 12 09:16:10--18:03:30 UT (MJD~59742.57$\pm$0.18), during which the telescope was in its 
6D, 1.5B, 6B configurations\footnote{\url{https://www.narrabri.atnf.csiro.au/operations/array_configurations/configurations.html}}, respectively. Data were recorded simultaneously at central frequencies of 5.5 and 9 GHz, with 2 GHz of bandwidth at each frequency band that was comprised of 2048 1-MHz channels. Data were first edited manually for radio frequency interference (RFI) before being calibrated following standard procedures within \textsc{casa}, version 5.2.1. We used J1939$-$6342 for bandpass and flux calibration, and B1600$-$489 for phase calibration (3.1$^\circ$ away). Imaging was carried out using the \textsc{casa} task \textsc{clean}, using a Briggs robust parameter of 0, balancing sensitivity and angular resolution. The radio counterpart of \blackhole\ was detected in all three observations at all frequency bands. We measured the flux density and position of the source using the \textsc{casa} task \textsc{imfit} by fitting the target with an elliptical Gaussian with the full width at half maximum (FWHM) values set to the parameters of the synthesized beam. Results are provided in supplementary information.

\subsection{X-ray observations}
\subsubsection{MAXI and Swift/BAT monitoring observations}
\label{sec:2_2_1}
The daily and orbital X-ray \textit{MAXI} light curves of the source are publicly available\footnote{\url{http://maxi.riken.jp/top/slist.html}}. We obtained the source's daily-averaged light curve in the 2--4~keV and 4--10~keV bands. We also extracted the 15--50 keV daily-averaged light curve from \textit{Swift/BAT}\footnote{\url{https://swift.gsfc.nasa.gov/results/transients/BlackHoles.html}}. See Figure~\ref{fig:MAXI_BAT_Swift_Nicer_MeerKAT_4U1543} for the monitoring light curve.

\subsubsection{Swift/XRT observations}
\label{sec:2_2_2}
There are 31 \textit{XRT} observations for \blackhole~during the 2021--2023 outburst, one of which (observation ID of 00014374020) is invalid with exposure time of 0. We extracted one energy spectrum per observation using the UK \textit{Swift} Science Data Centre pipeline, which provides quality spectra \citep{2007A&A_Evans,2009MNRAS_Evans} except one spectrum file (with observation ID of 00089352003; window timing mode) couldn't be generated. For this observation, we manually extracted source and background spectra, set up mkf file and generated arf using \textsc{xrtmkarf} task. We fitted each spectrum within \textsc{xspec} \citep{1996ASPC_Arnaud} with either a power-law/disc blackbody (power-law/diskbb in \textsc{xspec}) model or a combination of both models, modified by interstellar absorption. Therein we froze the Galactic neutral hydrogen absorption column density to the averaged value, $\rm N_H=0.338\times{10^{22}\,{\rm  cm}^{-2}}$, which is the Galactic value in the direction of the source \citep{2016A&A_HI4PI}, across all three spectrum files. The unabsorbed fluxes in the 1--10 keV band were extracted with the task \textsc{cflux}. See supplementary information 
for flux measurements and Figure~\ref{fig:MAXI_BAT_Swift_Nicer_MeerKAT_4U1543} for the light curve.

\subsubsection{NICER observations}
\label{sec:2_2_3}
\blackhole~was monitored densely by \textit{NICER} during its 2021--2023 outburst since 2021 June 12, one day after the beginning of its 2021--2023 X-ray outburst. We reduced the \textit{NICER} data using the pipeline tool \textsc{nicerl2} with the standard filtering and using ftool \textsc{xselect} to extract spectra. The background was estimated using the tool \textsc{nibackgen3C50}. The Focal Plane Module (FPM) No. 14 and No. 34 were removed because of the increased detector noise. The response matrix files (RMF) and ancillary files (ARF) were generated with the ftool \textsc{nicerrmf} and \textsc{nicerarf}. Similar to \textit{XRT} data reduction, we fitted the spectra within \textsc{xspec} using the same model recipe: considering the combination of a power-law and a diskbb model or only one component, modified by interstellar absorption. The Galactic neutral hydrogen absorption column density were fitted to $\rm N_H=0.338\times{10^{22}\,{\rm  cm}^{-2}}$ throughout the fit as what we did for \textit{XRT} data. The unabsorbed fluxes in the 1--10 keV band were extracted with \textsc{cflux}, the same task as we used for \textit{XRT} data. See supplementary information 
for flux measurements and Figure~\ref{fig:MAXI_BAT_Swift_Nicer_MeerKAT_4U1543} for the light curve.

\section{Results}
\subsection{The evolution of the radio and X-ray emission}
\label{sec:3_1}
Long-term X-ray monitoring observations show \blackhole~made a very quick hard-to-soft spectral state transition soon after it was found in its new 2021--2023 outburst occurred on 2021 June 11 (MJD 59376), as indicated by the rapid drop in hardness ratio (\textit{Swift/BAT}~15--50~keV/\textit{MAXI}~2--10~keV), which is shown in Figure~\ref{fig:MAXI_BAT_Swift_Nicer_MeerKAT_4U1543}. Subsequently, it rose in X-ray flux especially in soft X-ray bands dramatically, reaching a flux of 3~Crab in 2--4~keV at 22:51 UT on the same day \citep{2021ATel14701....1N} as seen by \textit{MAXI} ($\sim$~1~Crab in 2--10~keV on the same day), and the energy spectrum obtained from the four scan transits from 18:12 to 22:51 UT is well represented with an absorbed disc blackbody model with an inner disc temperature of $\sim$1~keV. Then the X-ray fluxes in all bands of \textit{MAXI} were continuing to increase, at 06:37 UT on June 14 (MJD 59379) the source reached 2--4~keV flux of 24.0$\pm$0.4 photons ${\rm cm}^{-2}\, {\rm s}^{-1}$ ($\sim$11.0 Crab; peaked at $\sim$~19 Crab in 2--10~keV on the same day as shown in Figure~\ref{fig:MAXI_BAT_Swift_Nicer_MeerKAT_4U1543}) and around that time the bolometric luminosity was $\sim$~1.34$\times$10$^{39}$ ergs {s}$^{-1}$, which is comparable to its Eddington luminosity (assuming the mass is 9.4 M$_{\odot}$), $\sim$~1.18$\times$10$^{39}$ ergs {s}$^{-1}$ \citep{2021ATel14708....1N}. \textit{NICER} started to monitor the source on June 12, just less than one day after the discovery of this outburst. 
The source had been in the soft state mostly since the discovery until the end of 2021, as evidenced by the fact that the hardness ratio (HR) obtained with \textit{MAXI} (4--10 keV flux over 2--4 keV flux) evolved stably and never surpassed 1 (see Figure~\ref{fig:MAXI_BAT_Swift_Nicer_MeerKAT_4U1543}), \textit{NICER} had been almost detecting only a disc component, and no core radio jet emission was confidently observed in our MeerKAT monitoring observations. 

The disc-dominated soft state was probably interrupted once -- on MJD 59453 when the source spectrum was found dominated by a disc with temperature of 0.9~keV with the requirement of an additional power-law component with a photon index of 2.6 that accounts for $\gtrsim$ 15\% of the total flux, suggestive of an intermediate state and a fast soft-intermediate-soft state transition occurring; in addition, we also noticed that around the state transition, there were active radio flaring events occurring from the core position \citep[][]{2021ATel14878....1Z}, these radio flaring activities reminiscent of those radio flares often observed during hard-to-soft state transitions in BH XRBs \citep[e.g.,][]{2002MNRAS.331..765B}. Notably, we discovered a transient jet ejection from the BH a couple of weeks after the state transition and found that the ejection moved apparent superluminally with intrinsic jet speed much higher than that of other Galactic BH transients. The detailed study of the superluminal ejections is detailed in a separate paper (Zhang et al. 2024, to submit). 

\blackhole~showed decreased X-ray flux in the following half year or so as seen by \textit{MAXI} and \textit{Swift}. A soft-to-hard state transition occurred, as suggested by the sudden increase in the HR on 2021 December 28 (MJD 59576) and the detection of a radio component with the flux density of 0.24$\pm$0.02 mJy with radio in-band spectral index of 0.28$\pm$0.49 at the Gaia optical position of the source \citep{2022ATel15157....1Z}. Since then, the source continued to decrease in its X-ray fluxes in the hard state and probably entered into its quiescent state on 2022 January 16 (MJD 59595), when its X-ray emission was modeled by a power-law with the photon index of $2.20^{+0.26}_{-0.24}$ and flux of $\sim$~3.30$\times$10$^{33}$ ergs {s}$^{-1}$ (at a distance of 5 kpc), and radio non-detection with 3$\sigma$~upper limit of 60~$\mu$Jy on the same day, consistent with a quiescent state. The source stayed in such a quiescent state until $\sim$~one month later, on February 14 (MJD 59624), when the source recovered its X-ray activities with a detectable X-ray flux with subsequent detection of the core radio jet emission one day later. From then on, the source started to evolve in a re-flaring phase in the hard state (five re-flares are found in total, not only in radio but also in X-rays), as shown in later stage of the outburst shown in Figure~\ref{fig:MAXI_BAT_Swift_Nicer_MeerKAT_4U1543}.

\subsection{Clues to a variable Lorentz factor of the compact jets}
\label{sec:3_2}
\begin{figure}
	\includegraphics[width=\columnwidth]{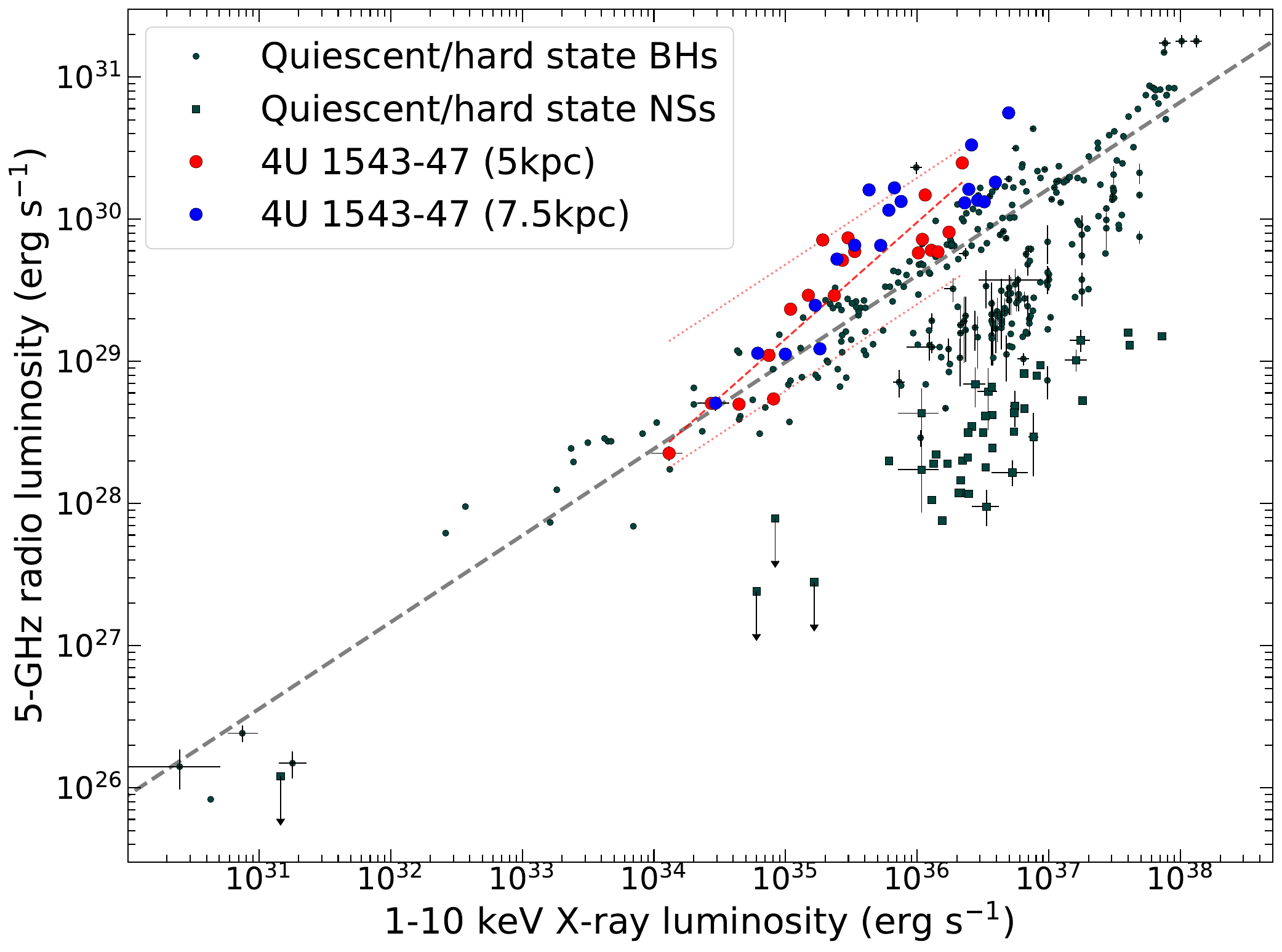}
    \caption{The radio/X-ray luminosity behaviours for a sample of BH XRBs (black filled circles) and NS XRBs (black filled squares) as collected in \citet{2018zenodo_Arashbahramian}. The gray dashed line shows the power-law fit ( $L_R\propto L_X^{0.61}$) for `radio-loud' track quiescent/hard state BHs collected in \citet{2018zenodo_Arashbahramian}. The hard state radio and X-ray luminosity measurements of \blackhole~are overplotted, using distances of 5 kpc and 7.5 kpc marked by the red and blue filled circles respectively. The \blackhole~is clearly radio brighter than other BHs and NSs; 
    we found a power-law correlation index between radio and X-ray luminosity of 0.82$\pm$0.09, as shown with the red dashed line. The highest and lowest nomalisations of the power-law relation (assuming the same index of 0.61) are shown by the two parallel red dashed lines, which could be explained by the variation of Lorentz factor of the hard state compact jet for~\blackhole, as detailed in Section~\ref{sec:3_2}.} 
    \label{fig:Radio-Xray_4U1543_2021-2022}
\end{figure}

Before our campaign, there was only one simultaneous radio/X-ray measurement obtained during its 2002 outburst, showing that the source is `radio bright' when assuming a source distance of 9~kpc \citep{Gallo_2003}. In the 2021--2023 X-ray outburst, 
we have collected 19 quasi-simultaneous radio and X-ray observations (within one day and chose the closest \textit{NICER} or \textit{Swift/XRT} observations in time with MeerKAT observations). In Figure~\ref{fig:Radio-Xray_4U1543_2021-2022}, we plot these measurements of the \blackhole~together with the radio/X-ray database of all X-ray binaries collected by \citet{2018zenodo_Arashbahramian}. Making use of the updated source distance of 5~kpc, which is obtained from combining the parallax measurement listed in Gaia Data Release 3 and assuming a Milky Way density prior for low-mass X-ray binaries (see also Section~\ref{sec:1_1}), we convert the 1.28 GHz radio flux densities to 5 GHz luminosities using the in-band spectral indices measured from corresponding MeerKAT epochs, while the 1--10 keV luminosities were estimated based on the 1--10 keV unabsorbed X-ray fluxes. We also apply the calculations to the case in which the distance is assumed to be 7.5~kpc, which is obtained from \citet{2002AAS_Orosz}. Both of these are shown in Figure~\ref{fig:Radio-Xray_4U1543_2021-2022}. We add 0.3 dex \citep[e.g.,][]{2021ApJ...907...34S} on measured errors for both radio and X-ray luminosities, to account for deviations from simultaneous measurements in the observations, the systematic uncertainties on using spectral index for different epochs of observations and systematics among instruments (e.g., \textit{Swift/XRT} and \textit{NICER}) used. We fit the quasi-simultaneous measurements using the Orthogonal Distance Regression (ODR) method packed in \textsc{scipy.odr}\footnote{\url{https://docs.scipy.org/doc/scipy/reference/odr.html}} under the formalism of $log L_R = \beta (log L_X-34) + N_R$ ($N_R$ represents radio normalization parameter), taking into account the errors on measured fluxes of both radio and X-ray observations, in which statistical errors on X-ray measurements have been chosen to be consistent (take the largest one from the left and right errors) before the fitting. As shown with red dashed line in Figure~\ref{fig:Radio-Xray_4U1543_2021-2022}, the fitted correlation line has a power law index of 0.82$\pm$0.09, which is steeper than the correlation index seen in other black holes as a whole \citep[e.g., $L_R\propto L_X^{0.61}$;][]{Gallo_2014}. The steeper slope hints possible dependence of the radio luminosity of the compact jet on radiative efficiency of the accretion flow. In black hole hard state, X-ray luminosity can be expressed as $L_{X} \propto \dot{M}^{q}$ where $q$ represents a radiatively efficient accretion flow ($q$=1) or a radiatively inefficient accretion flow ($q$=2--3) and $\dot{M}$ is mass accretion rate. In the framework of a coupled disc-jet scenario, the radio luminosity scales with the total jet power \citep[e.g.,][]{2003MNRAS.343L..59H} and mass accretion rate ($L_{\rm radio} \propto  \dot{M}^{\xi}$, where $\xi = (2p-(p+6)\alpha+13)/2(p+4)$, which depends on the spectrum index $\alpha$ of the jet and power-law index $p$ of the energy distribution of relativistic electrons). We can therefore write the relation between radio and X-ray luminosities as $L_{\rm radio} \propto  L_{X}^{\xi/q}$. The average spectral index of the compact jet of 4U 1543$-$47 is calculated as $\alpha\sim$ -0.24 while the classical values for $p$ is 2$<p<$3, producing $\xi$ with values in the range of 1.51--1.57, which correspond to values for $q=\xi/\beta$ in the range of 1.84$\pm$0.20--1.92$\pm$0.21. 
Both values for $q$ are close to a radiatively inefficient accretion flow ($q$=2--3) that is typical for a canonical black hole X-ray binary with a power-law radio/X-ray correlation index of around 0.6. 
Remarkably, the radio luminosities across a wide range of X-ray luminosities can be rather higher (i.e., `radio bright') as compared to those of the BH XRB sample, from $\sim$~10$^{34}$ ergs\,{s}$^{-1}$ to $\sim$~10$^{37}$ ergs\,{s}$^{-1}$. We discuss in detail two possibilities of the observed `radio bright' behaviour below.

There are two possible origins of the observed `radio bright' behaviour of the compact radio jet emission. The first one is that the observed core radio emission in hard state were contaminated by additional radio ejection component(s) that could not be resolved by MeerKAT observations. In order to investigate whether there are such component(s) dominating the radio emission we observed in the hard state, we observed the source with ATCA for three epochs, which provided us with high resolution capability to distinguish potential additional radio components. We could not detect any additional component(s) at these ATCA epochs. In addition, the spectral indices were rather flat. These suggest that the radio emission we measured very likely came from the core emission alone. The second possible origin is the radio bright behaviours are due to a variable Lorentz factor of the compact jets. To put a constraint on the range of the variable Lorentz factor, for each radio and X-ray simultaneous or quasi-simultaneous data pair in the radio/X-ray correlation plane, we fit a relation in the form of $log L_R = 0.61 (log L_X-34) + N_R$ to investigate the `radio brightness' by considering the free parameter $N_R$, while fix the index of the non-linear radio/X-ray luminosity correlation to a value of 0.61 that is typical for black hole transients. The radio/X-ray luminosity pair with the lowest normalization $N_R$ is $\sim$~28.182$\pm$0.040, probably indicative of the intrinsic luminosity of the \blackhole~in hard state; the highest normalization $N_R$ reached 29.072$\pm$0.004, differs from the lowest by $\sim$~0.9 dex (both normalizations are shown with two parallel red dotted lines in Figure~\ref{fig:Radio-Xray_4U1543_2021-2022}). This represents one of the largest spread in the radio normalization in a single source among black hole transients, suggestive of a larger dynamical range of the radio flux from the compact jet in \blackhole. 

The observed boosted radio luminosity and the intrinsic radio luminosity are related by $L_{obs} = L_{int}D^{n-\alpha_{s}}$, in which $\alpha_{s}$~is the spectral index of the radio emission from the compact core jet, $D$ is the Doppler factor that relates to Lorentz factor of $\Gamma=(1-(v/c)^{2})^{-1/2}$ ($v$ is the intrinsic speed of the jet making an angle of $\theta$~with the line of sight) as $D=\Gamma^{-1}[1-(v/c)cos\theta]^{-1}$, and here for a compact continuously replenished jet component we have $n = 2$ \citep{2006csxs.book..381F}. Thus the pair with the highest normalization (i.e., boosted most under the form of $D^{n-\alpha_{s}}$) has a Doppler factor of $\sim$~3 with Lorentz factor of $\sim$~1.9 ($\sim$~0.85 $c$; corresponds to the highest Lorentz factor of all pairs) under the assumption of a low jet inclination angle, e.g., 15$^{\circ}$ (the maximum allowed jet inclination angle is $\sim$~19.7$^{\circ}$). Alternatively, a varying inclination angle of the compact jet could produce the scatters in radio normalization, which might corresponds to the precession of the compact jet. Both factors discussed above can play their roles simultaneously to produce the large spread in the radio luminosity as observed in a relatively narrow range of X-ray luminosity as well as certain radio flux dependence on the X-ray luminosity in a wide range. 

\section{Conclusions}

\blackhole~went into its 2021--2023 outburst on 2021 June 11. The source made a fast hard-to-soft spectral state transition within one day, followed with a rise of its 2--10 keV X-ray flux significantly by $\sim$~4~orders of magnitude over a few days and then peaked at around 10$^{4}$ mCrab, which is likely close to its Eddington luminosity. Since then the source stayed almost entirely in the soft state until the end of 2021 when it took a short excursion of probably several days to the intermediate state. The source went back to the hard state and quiescent state at the end of 2021 and the beginning of 2022. Since then after remaining in quiescence for a month, it returned back to the hard state with multiple radio and X-ray re-brightening flares. The source shows a peculiar radio and X-ray behaviour during the hard state in terms of the radio and X-ray power-law correlation index, of 0.82$\pm$0.09, which is steeper than those of other BH XRBs; the source is also `radio brighter' than other BH and NS XRBs across several orders of magnitude in X-ray luminosities. The radio normalizations of all the quasi-simultaneously measured radio/X-ray data pairs differ by $\sim$~0.9 dex assuming the power-law index of 0.61, showing a large scatter among black hole transients. Since the source is found with a nearly face-on jet inclination angle from the measurements of proper motion of discrete jet ejections (detailed in a separate paper Zhang et al. 2024), the radio bright behaviour and large scattering in the radio luminosity of the radio emission can be explained by a variable Lorentz factor, in the range between 1 and $\sim$ 2, of the compact jet emission with a small jet inclination angle of \blackhole. 

\section*{Acknowledgements}
We thank Dr. Evangelia Tremou, Dr. Joe Bright, Dr. Andrew Hughes in the ThunderKAT collaboration for preparing the observing blocks and the staff at the South African Radio Astronomy Observatory for their rapid scheduling of these observations. The MeerKAT telescope is operated by the South African Radio Astronomy Observatory, which is a facility of the National Research Foundation, an agency of the Department of Science and Innovation. We thank Jamie Stevens and ATCA staff for scheduling the observations. We thank RIKEN, JAXA and the \textit{MAXI} team for making the \textit{MAXI} data available. We acknowledge the \textit{Swift} Guest Observing Facility for making \textit{BAT} data products available and the UK \textit{Swift} Science Data Centre for building most of the \textit{XRT} products in this research. We thank the \textit{NICER} team for making the data available. We acknowledge the use of the ilifu cloud computing facility -- www.ilifu.ac.za, a partnership between the University of Cape Town, the University of the Western Cape, Stellenbosch University, Sol Plaatje University, the Cape Peninsula University of Technology and the South African Radio Astronomy Observatory. The ilifu facility is supported by contributions from the Inter-University Institute for Data Intensive Astronomy (IDIA - a partnership between the University of Cape Town, the University of Pretoria and the University of the Western Cape), the Computational Biology division at UCT and the Data Intensive Research Initiative of South Africa (DIRISA). WY and XZ acknowledge support from the Natural Science Foundation of China (No. U1838203,12373050). FC acknowledges support from the Royal Society through the Newton International Fellowship programme (NIF/R1/211296).

\section*{Data Availability}
All raw data are publicly available. The corresponding reduced data and images in this work can only be shared with reasonable request.



\bibliographystyle{mnras}
\bibliography{4U1543} 




\section*{Online supplementary information}

\textbf{Table 1}. Information of the Radio (MeerKAT and ATCA) observations of the BH \blackhole~during its 2021--2023 outburst. Here we report on the flux measurements of the source whose positions are consistent with the core position. Uncertainties are quoted at the 1$\sigma$ level.

\noindent
\textbf{Table 2}. Information of the X-ray pointed (\textit{NICER} and \textit{Swift/XRT}) observations of the BH \blackhole~during its 2021--2023 outburst. Uncertainties are quoted at the 90\% level.


\onecolumn
\textbf{\large Online Supplementary Information}


\begin{center}
\begin{longtable}{*{5}{c}}
\caption {Information of the Radio (MeerKAT and ATCA) observations of the BH \blackhole~during its 2021--2023 outburst. Here we report on the flux measurements of the source whose positions are consistent with the core position. Uncertainties are quoted at the 1$\sigma$ level.}
\label{tab:rdata} \\

\hline\hline
\multicolumn{1}{c}{Epoch} & \multicolumn{1}{c}{MJD}  & \multicolumn{1}{c}{Telescope} & \multicolumn{1}{c}{Central frequency} & \multicolumn{1}{c}{$S_{\nu}$} \\

\multicolumn{1}{c}{} & \multicolumn{1}{c}{} & \multicolumn{1}{c}{} & \multicolumn{1}{c}{(GHz)} & \multicolumn{1}{c}{(mJy)} \\ \hline

\endfirsthead

\multicolumn{5}{c}
{{\tablename\ \thetable{} -- Continued from previous page.}} \\

\hline\hline 

\multicolumn{1}{c}{Epoch} & \multicolumn{1}{c}{MJD}  & \multicolumn{1}{c}{Telescope} & \multicolumn{1}{c}{Central frequency} & \multicolumn{1}{c}{$S_{\nu}$} \\

\multicolumn{1}{c}{} & \multicolumn{1}{c}{} & \multicolumn{1}{c}{} & \multicolumn{1}{c}{(GHz)} & \multicolumn{1}{c}{(mJy)}  \\ \hline

\endhead

\hline
\multicolumn{5}{c}{{Continued on next page}} \\ \hline
\endfoot

\hline
\endlastfoot

2021 Jun 19 & 59384.8843 & MeerKAT & 1.28 & 0.405$\pm$0.020 \\
 \hline
2021 Jun 27 & 59392.0078 & MeerKAT & 1.28 & 0.250$\pm$0.033 \\ 
 \hline
2021 Jul 04 & 59399.8759 & MeerKAT & 1.28 & 0.149$\pm$0.021 \\ 
 \hline
2021 Jul 11 & 59406.9940 & MeerKAT & 1.28 & 0.121$\pm$0.025 \\ 
 \hline
2021 Jul 26 & 59421.6770  & MeerKAT & 1.28 & $<$~0.066 \\ 
 \hline
2021 Jul 31 & 59426.8178 & MeerKAT & 1.28 & $<$~0.060 \\ 
 \hline
2021 Aug 07 & 59433.8102 & MeerKAT & 1.28 & $<$~0.060 \\ 
 \hline
2021 Aug 15 & 59441.8682 & MeerKAT & 1.28 & $<$~0.123 \\ 
 \hline
2021 Aug 22 & 59448.6218  & MeerKAT & 1.28 & 14.046$\pm$0.024 \\ 
 \hline
2021 Aug 28 & 59454.6255  & MeerKAT & 1.28 & 4.016$\pm$0.022 \\ 
 \hline
2021 Sep 05 & 59462.6971 & MeerKAT & 1.28 & $<$~0.054 \\ 
 \hline
2021 Sep 13 & 59470.5948 & MeerKAT & 1.28 & 13.320$\pm$0.025 \\ 
 \hline
2021 Sep 20 & 59477.6122 & MeerKAT & 1.28 & 0.170$\pm$0.022 \\ 
 \hline
2021 Sep 27 & 59484.5291  & MeerKAT & 1.28 & 0.218$\pm$0.023 \\ 
 \hline
2021 Oct 04 & 59491.5916  & MeerKAT & 1.28 & 0.090$\pm$0.022 \\ 
 \hline
2021 Oct 09 & 59496.5915 & MeerKAT & 1.28 & $<$~0.066 \\ 
 \hline
2021 Oct 17 & 59504.4557 & MeerKAT & 1.28 & $<$~0.057\\ 
 \hline
2021 Oct 23 & 59510.4661 & MeerKAT & 1.28 & $<$~0.057 \\ 
 \hline
2021 Oct 31 & 59518.5498  & MeerKAT & 1.28 & $<$~0.054 \\ 
 \hline
2021 Nov 08 & 59526.6119  & MeerKAT & 1.28 & $<$~0.069 \\ 
 \hline
2021 Nov 14 & 59532.4525 & MeerKAT & 1.28 & $<$~0.054 \\ 
\hline
2021 Nov 27 & 59545.3558 & MeerKAT & 1.28 & $<$~0.063 \\
\hline
2021 Dec 11 & 59559.5363 & MeerKAT & 1.28 & $<$~0.069 \\
\hline
2021 Dec 18 & 59566.4039 & MeerKAT & 1.28 & $<$~0.084 \\
\hline
2021 Dec 24 & 59572.4420 & MeerKAT & 1.28 & $<$~0.060 \\
\hline
2022 Jan 03 & 59582.3415 & MeerKAT & 1.28 & 0.242$\pm$0.019\\
\hline
2022 Jan 08 & 59587.4731 & MeerKAT & 1.28 & 0.330$\pm$0.025\\
\hline
2022 Jan 16 & 59595.2894 & MeerKAT & 1.28 & $<$~0.063 \\
\hline
2022 Jan 23 & 59602.2465 & MeerKAT & 1.28 & $<$~0.063 \\
\hline
2022 Jan 29 & 59608.4073 & MeerKAT & 1.28 & $<$~0.060 \\
\hline
2022 Feb 06 & 59616.3411 & MeerKAT & 1.28 & $<$~0.063 \\
\hline
2022 Feb 14 & 59624.1955 & MeerKAT & 1.28 & 4.538$\pm$0.021 \\
\hline
2022 Feb 18 & 59628.3620 & MeerKAT & 1.28 & 9.574$\pm$0.030 \\
\hline
2022 Feb 27 & 59637.3342 & MeerKAT & 1.28 & 16.386$\pm$0.028 \\
\hline
2022 Mar 07 & 59645.2300 & MeerKAT & 1.28 & 5.895$\pm$0.020 \\
\hline
2022 Mar 16 & 59654.2139 & MeerKAT & 1.28 & 5.832$\pm$0.023 \\
\hline
2022 Mar 19 & 59657.0635 & MeerKAT & 1.28 & 6.872$\pm$0.026 \\
\hline
2022 Mar 28 & 59666.0530 & MeerKAT & 1.28 & 6.871$\pm$0.022 \\
\hline
2022 Apr 04 & 59673.0310 & MeerKAT & 1.28 & 4.111$\pm$0.020 \\
\hline
2022 Apr 10 & 59679.0338 & MeerKAT & 1.28 & 2.080$\pm$0.022 \\
\hline
2022 Apr 16 & 59685.1686 & MeerKAT & 1.28 & 2.145$\pm$0.021 \\
\hline
2022 Apr 25 & 59694.0495 & MeerKAT & 1.28 & 4.835$\pm$0.017 \\
\hline
2022 Apr 27 & 59696.74 & ATCA & 5.5 & 4.48$\pm$0.03 \\
 & & & 9.0 & 4.09$\pm$0.02 \\
\hline
2022 Apr 30 & 59699.9899 & MeerKAT & 1.28 & 8.685$\pm$0.024 \\
\hline
2022 May 06 & 59705.9975 & MeerKAT & 1.28 & 7.413$\pm$0.021 \\
\hline
2022 May 15 & 59714.9959 & MeerKAT & 1.28 & 2.258$\pm$0.023 \\
\hline
2022 May 23 & 59722.8836 & MeerKAT & 1.28 & 0.625$\pm$0.022 \\
\hline
2022 May 28 & 59727.0830 & MeerKAT & 1.28 & 0.472$\pm$0.023 \\
\hline
2022 May 28 & 59727.48 & ATCA & 5.5 & 1.14$\pm$0.02 \\
 & & & 9.0 & 1.03$\pm$0.02 \\
\hline
2022 Jun 03 & 59733.8830 & MeerKAT & 1.28 & 5.233$\pm$0.021 \\
\hline
2022 Jun 12 & 59742.9142 & MeerKAT & 1.28 & 29.886$\pm$0.024 \\
\hline
2022 Jun 12 & 59742.39 & ATCA & 5.5 & 5.95$\pm$0.06 \\
 & & & 9.0 & 4.68$\pm$0.04 \\
\hline
2022 Jun 17 & 59747.8624 & MeerKAT & 1.28 & 5.285$\pm$0.020 \\
\hline
2022 Jun 29 & 59759.9280 & MeerKAT & 1.28 & 1.833$\pm$0.021 \\
\hline
2022 Jul 03 & 59763.8626 & MeerKAT & 1.28 & 1.436$\pm$0.020 \\
\hline
2022 Jul 09 & 59769.8164 & MeerKAT & 1.28 & 0.204$\pm$0.018 \\
\hline
2022 Jul 15 & 59775.7679 & MeerKAT & 1.28 & $<$~0.069 \\
\hline
2022 Jul 22 & 59782.8094 & MeerKAT & 1.28 & $<$~0.069 \\
\hline
2022 Oct 28 & 59880.6445 & MeerKAT & 1.28 & 16.032$\pm$0.026 \\
\hline
2022 Nov 06 & 59889.6495 & MeerKAT & 1.28 & 7.012$\pm$0.025 \\
\hline
2022 Nov 11 & 59894.6079 & MeerKAT & 1.28 & 6.249$\pm$0.022 \\
\hline
2022 Nov 18 & 59901.5453 & MeerKAT & 1.28 & 3.033$\pm$0.024 \\
\hline
2022 Nov 26 & 59909.5446 & MeerKAT & 1.28 & 2.611$\pm$0.025 \\
\hline
2022 Dec 02 & 59915.5835 & MeerKAT & 1.28 & 1.382$\pm$0.040 \\
\hline
2022 Dec 09 & 59922.5454 & MeerKAT & 1.28 & $<$~0.090 \\
\hline
2022 Dec 19 & 59932.3174 & MeerKAT & 1.28 & $<$~0.084 \\
\hline
2022 Dec 26 & 59939.2746 & MeerKAT & 1.28 & $<$~0.081 \\
\hline
2023 Jan 02 & 59946.4028 & MeerKAT & 1.28 & $<$~0.072 \\
\hline 
2023 Jan 07 & 59951.3464 & MeerKAT & 1.28 & $<$~0.072 \\
\hline
2023 Jan 14 & 59958.3514 & MeerKAT & 1.28 & $<$~0.099 \\
\hline
2023 Jan 21 & 59965.3888 & MeerKAT & 1.28 & $<$~0.060 \\
\hline
2023 Jan 28 & 59972.3107 & MeerKAT & 1.28 & $<$~0.066 \\
\hline
2023 Feb 04 & 59979.2122 & MeerKAT & 1.28 & $<$~0.069 \\
\hline
2023 Feb 21 & 59996.1940 & MeerKAT & 1.28 & $<$~0.060 \\
\hline
2023 Mar 09 & 60012.3161 & MeerKAT & 1.28 & $<$~0.072 \\

\end{longtable}

\end{center}


\begin{center}
\begin{longtable}{*{9}{c}}
\caption {Information of the X-ray pointed (\textit{NICER} and \textit{Swift/XRT}) observations of the BH \blackhole~during its 2021--2023 outburst. Uncertainties are quoted at the 90\% level.}
\label{tab:xdata} \\

\hline\hline
\multicolumn{1}{c}{Epoch} & \multicolumn{1}{c}{MJD} & \multicolumn{1}{c}{Telescope} & \multicolumn{1}{c}{Exposure Time} & \multicolumn{1}{c}{Obs. ID} & \multicolumn{1}{c}{{$F_{unabs}$}} & \multicolumn{1}{c}{$kT$} & \multicolumn{1}{c}{{$\Gamma$}} & \multicolumn{1}{c}{${\chi^2}/(dof)$}\\ 

\multicolumn{1}{c}{} & \multicolumn{1}{c}{} & \multicolumn{1}{c}{} & \multicolumn{1}{c}{(seconds)} & \multicolumn{1}{c}{} & \multicolumn{1}{c}{({10$^{-10}${erg\,s}$^{-1}${cm}$^{-2}$})} & \multicolumn{1}{c}{(keV)} & \multicolumn{1}{c}{} & \\ [0.15cm]
\hline

\endfirsthead

\multicolumn{9}{c}
{{\tablename\ \thetable{} -- Continued from previous page. 
}} \\

\hline\hline 

\multicolumn{1}{c}{Epoch} & \multicolumn{1}{c}{MJD} & \multicolumn{1}{c}{Telescope} & \multicolumn{1}{c}{Exposure Time} & \multicolumn{1}{c}{Obs. ID} & \multicolumn{1}{c}{{$F_{unabs}$}} & \multicolumn{1}{c}{kT} & \multicolumn{1}{c}{{$\Gamma$}} & \multicolumn{1}{c}{${\chi^2}/(dof)$}\\

\multicolumn{1}{c}{} & \multicolumn{1}{c}{} & \multicolumn{1}{c}{} & \multicolumn{1}{c}{(seconds)} & \multicolumn{1}{c}{} & \multicolumn{1}{c}{({10$^{-10}${erg\,s}$^{-1}${cm}$^{-2}$})} & \multicolumn{1}{c}{(keV)} & \multicolumn{1}{c}{} & \\ [0.15cm]
\hline

\endhead

\hline
\multicolumn{9}{c}{{Continued on next page}} \\ \hline
\endfoot

\hline
\endlastfoot

2021 Jun 12 & 59377.5303 & \textit{Swift/XRT} & 1176	& 01055000000 & \phantom{0}$1875^{+8}_{-9}$ & \phantom{0}$1.429^{+0.006}_{-0.006}$ & - & 1.77\\[0.20cm]
\hline

2021 Jun 13 & 59378.2691 & \textit{Swift/XRT} & 1924	& 00014374001 & \phantom{0}$2608^{+9}_{-10}$ & \phantom{0}$1.495^{+0.008}_{-0.008}$ & - & 1.63\\[0.20cm]
\hline

2021 Jun 14 & 59379.3310 & \textit{Swift/XRT} & 1939	& 00014374002 & \phantom{0}$3097^{+11}_{-12}$ & \phantom{0}$1.598^{+0.009}_{-0.009}$ & - & 1.60\\[0.20cm]
\hline

2021 Jun 15 & 59380.9435 & \textit{Swift/XRT} & 760 & 00014374004 & \phantom{0}$3260^{+20}_{-18}$ & \phantom{0}$1.596^{+0.013}_{-0.013}$ & - & 1.29\\[0.20cm]
\hline

2021 Jun 16 & 59381.7976 & \textit{Swift/XRT} & 945 & 00014374005 & \phantom{0}$3278^{+17}_{-17}$ & \phantom{0}$1.573^{+0.011}_{-0.012}$ & - & 1.27\\[0.20cm]
\hline

2021 Jun 24 & 59389.3796 & \textit{Swift/XRT} & 560 & 00014374006 & \phantom{0}$1414^{+9}_{-10}$ & \phantom{0}$1.375^{+0.012}_{-0.012}$ & - & 1.36\\[0.20cm]
\hline

2021 Jun 26 & 59391.5518 & \textit{Swift/XRT} & 519 & 00014374008 & \phantom{0}$1530^{+9}_{-10}$ & \phantom{0}$1.328^{+0.011}_{-0.011}$ & - & 1.24\\[0.20cm]
\hline

2021 Jun 27 & 59392.0768 & \textit{Swift/XRT} & 609 & 00014374009 & \phantom{0}$1618^{+10}_{-11}$ & \phantom{0}$1.331^{+0.011}_{-0.011}$ & - & 1.24\\[0.20cm]
\hline

2021 Jul 20 & 59415.1125 & \textit{NICER} & 1604	& 4202230137 & \phantom{0}$704^{+0.253}_{-0.253}$ & \phantom{0}$0.985^{+0.000437}_{-0.000435}$ & - & 1.82\\[0.20cm]
\hline

2021 Jul 21 & 59416.5891 & \textit{NICER} & 627	& 4202230138 & \phantom{0}$691^{+0.426}_{-0.337}$ & \phantom{0}$0.981^{+0.000686}_{-0.000649}$ & - & 1.51 \\[0.20cm]
\hline

2021 Jul 27 & 59422.0164 & \textit{Swift/XRT} & 1828 & 00089352001 & \phantom{0}$710.3^{+2.2}_{-2.3}$ & \phantom{0}$1.078^{+0.004}_{-0.004}$ & - & 1.42\\[0.20cm]
\hline

2021 Jul 30 & 59425.8029 & \textit{NICER} & 1809	& 4202230140 & \phantom{0}$655^{+0.297}_{-0.348}$ & \phantom{0}$0.955^{+0.000441}_{-0.000420}$ & - & 1.92 \\[0.20cm]
\hline

2021 Jul 31 & 59426.1273 & \textit{NICER} & 1301	& 4202230141 & \phantom{0}$655^{+0.374}_{-0.301}$ & \phantom{0}$0.956^{+0.000511}_{-0.000538}$ & - & 1.71 \\[0.20cm]
\hline

2021 Aug 01 & 59427.2257 & \textit{NICER} & 2158	& 4202230142 & \phantom{0}$652^{+0.294}_{-0.215}$ & \phantom{0}$0.956^{+0.000400}_{-0.000397}$ & - & 1.62 \\[0.20cm]
\hline

2021 Aug 02 & 59428.1298 & \textit{NICER} & 892	& 4202230143 & \phantom{0}$643^{+0.279}_{-0.318}$ & \phantom{0}$0.967^{+0.000640}_{-0.000641}$ & - & 1.78 \\[0.20cm]

& 59428.5233 & \textit{Swift/XRT} & 1699 & 00089352002 & \phantom{0}$675.3^{+2.2}_{-2.2}$ & \phantom{0}$1.056^{+0.004}_{-0.004}$ & - & 1.30\\[0.20cm]
\hline

2021 Aug 03 & 59429.0986 & \textit{NICER} & 430 & 4202230144 & \phantom{0}$631^{+0.593}_{-0.435}$ & \phantom{0}$0.950^{+0.000900}_{-0.000935}$ & - & 1.41 \\[0.20cm]
\hline

2021 Aug 04 & 59430.3083 & \textit{NICER} & 592 & 4202230145 & \phantom{0}$641^{+0.487}_{-0.378}$ & \phantom{0}$0.951^{+0.000756}_{-0.000789}$ & - & 1.19 \\[0.20cm]
\hline

2021 Aug 05 & 59431.0870 & \textit{NICER} & 745 & 4202230146 & \phantom{0}$625^{+0.364}_{-0.358}$ & \phantom{0}$0.946^{+0.000682}_{-0.000707}$ & - & 1.34 \\[0.20cm]
\hline

2021 Aug 06 & 59432.6477 & \textit{NICER} & 657 & 4202230147 & \phantom{0}$633^{+0.379}_{-0.466}$ & \phantom{0}$0.947^{+0.000721}_{-0.000728}$ & - & 1.16 \\[0.20cm]
\hline

2021 Aug 07 & 59433.1685 & \textit{NICER} & 769 & 4202230148 & \phantom{0}$632^{+0.287}_{-0.467}$ & \phantom{0}$0.950^{+0.000672}_{-0.000691}$ & - & 1.19 \\[0.20cm]
\hline

2021 Aug 10 & 59436.4470 & \textit{NICER} & 383 & 4202230149 & \phantom{0}$625^{+0.492}_{-0.640}$ & \phantom{0}$0.944^{+0.000968}_{-0.000944}$ & - & 1.25 \\[0.20cm]
\hline

2021 Aug 12 & 59438.2441 & \textit{NICER} & 114 & 4202230150 & \phantom{0}$620^{+1.150}_{-0.892}$ & \phantom{0}$0.936^{+0.00164}_{-0.00165}$ & - & 1.28 \\[0.20cm]
\hline

2021 Aug 17 & 59443.2938 & \textit{NICER} & 971 & 4202230151 & \phantom{0}$616^{+0.487}_{-0.356}$ & \phantom{0}$0.942^{+0.000606}_{-0.000592}$ & - & 1.43 \\[0.20cm]
\hline

2021 Aug 25 & 59451.9444 & \textit{NICER} & 27 & 4202230152 & \phantom{0}$595^{+1.690}_{-1.970}$ & \phantom{0}$0.932^{+0.00355}_{-0.00354}$ & - & 1.21 \\[0.20cm]
\hline

2021 Aug 26 & 59452.0087 & \textit{NICER} & 279 & 4202230153 & \phantom{0}$587^{+0.615}_{-0.647}$ & \phantom{0}$0.938^{+0.00140}_{-0.000954}$ & - & 0.99 \\[0.20cm]
\hline

2021 Aug 27 & 59453.6248 & \textit{NICER} & 401 & 4202230154 & \phantom{0}$592^{+0.927}_{-3.550}$ & \phantom{0}$0.903^{+0.00300}_{-0.00261}$ & \phantom{0}$2.58^{+0.14}_{-0.15}$ & 1.00 \\[0.20cm]
\hline

2021 Aug 28 & 59454.0123 & \textit{NICER} & 403 & 4202230155 & \phantom{0}$589^{+2.630}_{-8.750}$ & \phantom{0}$0.906^{+0.00346}_{-0.00319}$ & \phantom{0}$2.80^{+0.13}_{-0.15}$ & 0.98 \\[0.20cm]
\hline

2021 Aug 31 & 59457.3066 & \textit{NICER} & 578 & 4202230158 & \phantom{0}$569^{+0.326}_{-0.479}$ & \phantom{0}$0.932^{+0.000838}_{-0.000777}$ & - & 1.05 \\[0.20cm]
\hline

2021 Sep 02 & 59459.9993 & \textit{NICER} & 639 & 4202230161 & \phantom{0}$546^{+0.441}_{-0.589}$ & \phantom{0}$0.913^{+0.000760}_{-0.000833}$ & - & 1.32 \\[0.20cm]
\hline

2021 Sep 04 & 59461.0488 & \textit{NICER} & 1225 & 4202230162 & \phantom{0}$533^{+0.266}_{-0.307}$ & \phantom{0}$0.910^{+0.000551}_{-0.000536}$ & - & 1.41 \\[0.20cm]
\hline

2021 Sep 07 & 59464.1972 & \textit{NICER} & 635 & 4202230165 & \phantom{0}$529^{+0.489}_{-0.431}$ & \phantom{0}$0.913^{+0.000779}_{-0.000757}$ & - & 1.06 \\[0.20cm]
\hline

2021 Sep 09 & 59466.0695 & \textit{NICER} & 1722 & 4202230166 & \phantom{0}$517^{+0.235}_{-0.222}$ & \phantom{0}$0.909^{+0.000471}_{-0.000471}$ & - & 1.09 \\[0.20cm]
\hline

2021 Sep 10 & 59467.0382 & \textit{NICER} & 1323 & 4202230167 & \phantom{0}$512^{+0.243}_{-0.279}$ & \phantom{0}$0.903^{+0.000531}_{-0.000521}$ & - & 1.22 \\[0.20cm]
\hline

2021 Sep 11 & 59468.5875 & \textit{NICER} & 2528 & 4202230168 & \phantom{0}$521^{+0.146}_{-0.289}$ & \phantom{0}$0.918^{+0.000390}_{-0.000379}$ & - & 1.55 \\[0.20cm]
\hline

2021 Sep 12 & 59469.1053 & \textit{NICER} & 4391 & 4202230169 & \phantom{0}$526^{+1.560}_{-1.720}$ & \phantom{0}$0.913^{+0.000940}_{-0.000903}$ & \phantom{0}$3.13^{+0.08}_{-0.08}$ & 1.41 \\[0.20cm]
\hline

2021 Sep 13 & 59470.1370 & \textit{NICER} & 2619 & 4202230170 & \phantom{0}$507^{+0.854}_{-2.030}$ & \phantom{0}$0.892^{+0.000953}_{-0.000869}$ & \phantom{0}$2.68^{+0.11}_{-0.12}$ & 1.33 \\[0.20cm]
\hline

2021 Sep 14 & 59471.4324 & \textit{NICER} & 1022 & 4202230171 & \phantom{0}$508^{+0.370}_{-0.359}$ & \phantom{0}$0.911^{+0.000620}_{-0.000616}$ & - & 1.07 \\[0.20cm]

& 59471.5997 & \textit{Swift/XRT} & 1828 & 00089352004 & \phantom{0}$538.2^{+1.8}_{-1.7}$ & \phantom{0}$0.984^{+0.004}_{-0.004}$ & - & 1.28\\[0.20cm]
\hline

2021 Sep 15 & 59472.3465 & \textit{NICER} & 702 & 4202230172 & \phantom{0}$507^{+0.350}_{-0.367}$ & \phantom{0}$0.914^{+0.000730}_{-0.000731}$ & - & 1.14 \\[0.20cm]
\hline

2021 Sep 16 & 59473.4470 & \textit{NICER} & 401 & 4202230173 & \phantom{0}$492^{+0.479}_{-0.584}$ & \phantom{0}$0.898^{+0.000978}_{-0.000979}$ & - & 1.05 \\[0.20cm]
\hline

2021 Sep 18 & 59475.1125 & \textit{NICER} & 760 & 4202230175 & \phantom{0}$481^{+0.442}_{-0.410}$ & \phantom{0}$0.897^{+0.000725}_{-0.000723}$ & - & 1.25 \\[0.20cm]
\hline

2021 Sep 24 & 59481.5771 & \textit{NICER} & 496 & 4202230180 & \phantom{0}$459^{+0.405}_{-0.355}$ & \phantom{0}$0.900^{+0.000902}_{-0.000904}$ & - & 1.28 \\[0.20cm]
\hline

2021 Sep 27 & 59484.1491 & \textit{NICER} & 345 & 4202230183 & \phantom{0}$411^{+0.540}_{-0.430}$ & \phantom{0}$0.881^{+0.00114}_{-0.00112}$ & - & 1.02 \\[0.20cm]
\hline

2021 Sep 28 & 59485.1794 & \textit{NICER} & 780 & 4202230184 & \phantom{0}$402^{+0.370}_{-0.284}$ & \phantom{0}$0.870^{+0.000749}_{-0.000746}$ & - & 1.22 \\[0.20cm]
\hline

2021 Sep 30 & 59487.8928 & \textit{NICER} & 594 & 4202230185 & \phantom{0}$388^{+0.349}_{-0.405}$ & \phantom{0}$0.862^{+0.000858}_{-0.000854}$ & - & 1.13 \\[0.20cm]
\hline

2021 Oct 11 & 59498.7151 & \textit{NICER} & 1336 & 4202230187 & \phantom{0}$344^{+0.347}_{-0.230}$ & \phantom{0}$0.828^{+0.000585}_{-0.000584}$ & - & 1.26 \\[0.20cm]
\hline

2021 Oct 14 & 59501.3026 & \textit{NICER} & 598 & 4202230188 & \phantom{0}$329^{+0.492}_{-0.137}$ & \phantom{0}$0.821^{+0.000881}_{-0.000896}$ & - & 1.01 \\[0.20cm]
\hline

2022 Jan 08 & 59587.0403 & \textit{NICER} & 1171 & 4202230192 & \phantom{0}$0.0913^{+0.00422}_{-0.00444}$ & - & \phantom{0}$1.79^{+0.07}_{-0.07}$ & 1.38 \\[0.20cm]
\hline

2022 Jan 15 & 59594.1760 & \textit{Swift/XRT} & 938 & 00089352005 & \phantom{0}$0.00909^{+0.01251}_{-0.00586}$ & - & \phantom{0}$2.41^{+1.35}_{-1.12}$ & 1.08\\[0.20cm]
\hline

2022 Jan 16 & 59595.6715 & \textit{NICER} & 1087 & 4202230195 & \phantom{0}$0.0110^{+0.00221}_{-0.00133}$ & - & \phantom{0}$2.20^{+0.26}_{-0.24}$ & 1.56 \\[0.20cm]
\hline

2022 Feb 15 & 59625.1162 & \textit{NICER} & 1179 & 4202230206 & \phantom{0}$5.83^{+0.0151}_{-0.0352}$ & \phantom{0}$0.166^{+0.00682}_{-0.00664}$ & \phantom{0}$1.77^{+0.01}_{-0.01}$ & 1.07 \\[0.20cm]
\hline

2022 Feb 23 & 59633.3537 & \textit{NICER} & 550 & 4202230208 & \phantom{0}$9.79^{+0.0537}_{-0.0538}$ & - & \phantom{0}$1.64^{+0.01}_{-0.01}$ & 1.07 \\[0.20cm]
\hline

2022 Feb 24 & 59634.0000 & \textit{NICER} & 972 & 4202230209 & \phantom{0}$8.02^{+0.0450}_{-0.0367}$ & \phantom{0}$0.181^{+0.00691}_{-0.00670}$ & \phantom{0}$1.74^{+0.01}_{-0.01}$ & 1.14 \\[0.20cm]
\hline

2022 Feb 25 & 59635.0359 & \textit{NICER} & 2191 & 4202230210 & \phantom{0}$6.58^{+0.0228}_{-0.0228}$ & \phantom{0}$0.186^{+0.00544}_{-0.00555}$ & \phantom{0}$1.67^{+0.01}_{-0.01}$ & 1.35 \\[0.20cm]
\hline

2022 Feb 26 & 59636.3903 & \textit{NICER} & 3039 & 4681010101 & \phantom{0}$4.73^{+0.0133}_{-0.0192}$ & - & \phantom{0}$1.62^{+0.01}_{-0.01}$ & 1.15 \\[0.20cm]
\hline

2022 Feb 27 & 59637.0336 & \textit{NICER} & 8146 & 4681010102 & \phantom{0}$3.84^{+0.0113}_{-0.00689}$ & - & \phantom{0}$1.57^{+0.01}_{-0.01}$ & 1.15 \\[0.20cm]
\hline

2022 Feb 28 & 59638.0046 & \textit{NICER} & 6893 & 4681010103 & \phantom{0}$2.93^{+0.00924}_{-0.00723}$ & - & \phantom{0}$1.61^{+0.01}_{-0.01}$ & 1.08 \\[0.20cm]
\hline

2022 Mar 05 & 59643.0115 & \textit{NICER} & 3928 & 4681010104 & \phantom{0}$1.22^{+0.00592}_{-0.00535}$ & - & \phantom{0}$1.81^{+0.01}_{-0.01}$ & 1.16 \\[0.20cm]
\hline

2022 Mar 06 & 59644.0692 & \textit{Swift/XRT} & 1238 & 00089352011 & \phantom{0}$0.997^{+0.0534}_{-0.0528}$ & - & \phantom{0}$1.60^{+0.06}_{-0.06}$ & 0.96 \\[0.20cm]
\hline

2022 Mar 09 & 59647.3269 & \textit{NICER} & 1346 & 5530010101 & \phantom{0}$0.801^{+0.00644}_{-0.00825}$ & - & \phantom{0}$1.95^{+0.01}_{-0.01}$ & 1.20 \\[0.20cm]
\hline

2022 Mar 10 & 59648.1076 & \textit{NICER} & 2677 & 4681010201 & \phantom{0}$0.892^{+0.00661}_{-0.00651}$ & - & \phantom{0}$1.86^{+0.01}_{-0.01}$ & 1.00 \\[0.20cm]

 & 59648.3673 & \textit{NICER} & 620 & 5530010102 & \phantom{0}$0.829^{+0.0130}_{-0.0115}$ & - & \phantom{0}$1.96^{+0.02}_{-0.02}$ & 1.18 \\[0.20cm]
\hline

2022 Mar 11 & 59649.3296 & \textit{NICER} & 625 & 5530010103 & \phantom{0}$0.982^{+0.0156}_{-0.0171}$ & - & \phantom{0}$1.81^{+0.02}_{-0.02}$ & 0.92 \\[0.20cm]
\hline

2022 Mar 12 & 59650.3617 & \textit{NICER} & 673 & 5530010104 & \phantom{0}$1.390^{+0.0190}_{-0.0154}$ & - & \phantom{0}$1.80^{+0.02}_{-0.02}$ & 1.15 \\[0.20cm]
\hline

2022 Mar 13 & 59651.4316 & \textit{Swift/XRT} & 933 & 00089352012 & \phantom{0}$1.998^{+0.122}_{-0.121}$ & - & \phantom{0}$1.66^{+0.07}_{-0.07}$ & 1.06 \\[0.20cm]
\hline

2022 Mar 14 & 59652.9472 & \textit{NICER} & 566 & 4681010301 & \phantom{0}$2.54^{+0.0258}_{-0.0173}$ & - & \phantom{0}$1.91^{+0.01}_{-0.01}$ & 1.50 \\[0.20cm]
\hline

2022 Mar 15 & 59653.2704 & \textit{NICER} & 491 & 5530010105 & \phantom{0}$2.57^{+0.0222}_{-0.0199}$ & - & \phantom{0}$1.93^{+0.01}_{-0.01}$ & 1.24 \\[0.20cm]
\hline

2022 Mar 16 & 59654.2411 & \textit{NICER} & 455 & 5530010106 & \phantom{0}$3.41^{+0.0306}_{-0.0449}$ & \phantom{0}$0.207^{+0.0202}_{-0.0204}$ & \phantom{0}$1.61^{+0.04}_{-0.04}$ & 1.19 \\[0.20cm]
\hline

2022 Mar 19 & 59657.0778 & \textit{NICER} & 879 & 4681010401 & \phantom{0}$4.28^{+0.0220}_{-0.0294}$ & \phantom{0}$0.166^{+0.00967}_{-0.00931}$ & \phantom{0}$1.80^{+0.02}_{-0.02}$ & 1.07 \\[0.20cm]
\hline

2022 Mar 20 & 59658.3192 & \textit{Swift/XRT} & 728 & 00089352013 & \phantom{0}$4.605^{+0.278}_{-0.272}$ & - & \phantom{0}$1.86^{+0.07}_{-0.07}$ & 1.04 \\[0.20cm]
\hline

2022 Mar 22 & 59660.2857 & \textit{NICER} & 1795 & 5530010107 & \phantom{0}$6.45^{+0.0243}_{-0.0283}$ & - & \phantom{0}$1.67^{+0.01}_{-0.01}$ & 0.96 \\[0.20cm]
\hline

2022 Mar 24 & 59662.5442 & \textit{NICER} & 3346 & 5202230201 & \phantom{0}$6.08^{+0.0149}_{-0.0160}$ & - & \phantom{0}$1.63^{+0.01}_{-0.01}$ & 1.09 \\[0.20cm]
\hline

2022 Mar 26 & 59664.351 & \textit{NICER} & 1106 & 5530010108 & \phantom{0}$4.49^{+0.0222}_{-0.0329}$ & - & \phantom{0}$1.64^{+0.01}_{-0.01}$ & 0.93 \\[0.20cm]
\hline

2022 Mar 27 & 59665.2226 & \textit{Swift/XRT} & 980 & 00089352014 & \phantom{0}$3.657^{+0.198}_{-0.193}$ & - & \phantom{0}$1.66^{+0.06}_{-0.06}$ & 0.88 \\[0.20cm]
\hline

2022 Mar 31 & 59669.9634 & \textit{NICER} & 1110 & 5202230202 & \phantom{0}$0.894^{+0.0124}_{-0.0122}$ & - & \phantom{0}$1.88^{+0.01}_{-0.01}$ & 1.03 \\[0.20cm]
\hline

2022 Apr 01 & 59670.1574 & \textit{NICER} & 411 & 5202230203 & \phantom{0}$0.855^{+0.0123}_{-0.0141}$ & - & \phantom{0}$1.87^{+0.02}_{-0.02}$ & 1.03 \\[0.20cm]
\hline

2022 Apr 03 & 59672.7246 & \textit{Swift/XRT} & 758 & 00089352015 & \phantom{0}$0.792^{+0.098}_{-0.103}$ & - & \phantom{0}$1.76^{+0.15}_{-0.14}$ & 1.26 \\[0.20cm]

& 59672.8669 & \textit{NICER} & 1605 & 5202230204 & \phantom{0}$0.640^{+0.00794}_{-0.00669}$ & - & \phantom{0}$1.97^{+0.02}_{-0.02}$ & 1.03 \\[0.20cm]
\hline

2022 Apr 07 & 59676.9317 & \textit{NICER} & 574 & 5202230205 & \phantom{0}$0.575^{+0.0111}_{-0.0129}$ & - & \phantom{0}$1.96^{+0.03}_{-0.03}$ & 1.01 \\[0.20cm]

 & 59676.9960 & \textit{NICER} & 80 & 5202230206 & \phantom{0}$0.779^{+0.0406}_{-0.0375}$ & - & \phantom{0}$1.59^{+0.07}_{-0.07}$ & 1.48 \\[0.20cm]
\hline

2022 Apr 10 & 59679.0609 & \textit{NICER} & 1309 & 5202230207 & \phantom{0}$0.784^{+0.0158}_{-0.0139}$ & - & \phantom{0}$1.29^{+0.04}_{-0.04}$ & 1.42 \\[0.20cm]

& 59679.1024 & \textit{Swift/XRT} & 1033 & 00089352016 & \phantom{0}$0.500^{+0.0330}_{-0.0330}$ & - & \phantom{0}$1.89^{+0.12}_{-0.12}$ & 1.13 \\[0.20cm]
\hline

2022 Apr 13 & 59682.4954 & \textit{NICER} & 1591 & 5202230208 & \phantom{0}$0.448^{+0.00606}_{-0.00781}$ & - & \phantom{0}$2.04^{+0.02}_{-0.02}$ & 1.11 \\[0.20cm]
\hline

2022 Apr 17 & 59686.5351 & \textit{Swift/XRT} & 696 & 00014374010 & \phantom{0}$0.498^{+0.0399}_{-0.0387}$ & - & \phantom{0}$1.88^{+0.09}_{-0.09}$ & 0.93 \\[0.20cm]

& 59686.6748 & \textit{NICER} & 389 & 5202230209 & \phantom{0}$0.754^{+0.0210}_{-0.0177}$ & - & \phantom{0}$1.69^{+0.03}_{-0.03}$ & 1.60 \\[0.20cm]
\hline

2022 Apr 22 & 59691.8254 & \textit{Swift/XRT} & 952 & 00014374011 & \phantom{0}$0.667^{+0.0994}_{-0.0492}$ & - & \phantom{0}$1.70^{+0.09}_{-0.08}$ & 1.11 \\[0.20cm]
\hline

2022 Apr 24 & 59693.0796 & \textit{NICER} & 1469 & 5202230211 & \phantom{0}$0.904^{+0.00853}_{-0.00949}$ & - & \phantom{0}$1.86^{+0.01}_{-0.01}$ & 1.13 \\[0.20cm]
\hline

2022 May 01 & 59700.2917 & \textit{NICER} & 1571 & 5202230214 & \phantom{0}$4.800^{+0.0137}_{-0.0300}$ & - & \phantom{0}$1.77^{+0.01}_{-0.01}$ & 1.22 \\[0.20cm]
\hline

2022 May 05 & 59704.3562 & \textit{NICER} & 1054 & 5202230215 & \phantom{0}$2.400^{+0.0541}_{-0.0729}$ & - & \phantom{0}$1.74^{+0.02}_{-0.02}$ & 1.06 \\[0.20cm]
\hline

2022 May 07 & 59706.7469 & \textit{Swift/XRT} & 687 & 00014374012 & \phantom{0}$1.115^{+0.0770}_{-0.0750}$ & - & \phantom{0}$1.81^{+0.08}_{-0.08}$ & 1.04 \\[0.20cm]
\hline

2022 May 08 & 59707.5841 & \textit{NICER} & 1536 & 5202230216 & \phantom{0}$0.949^{+0.00879}_{-0.00880}$ & - & \phantom{0}$1.83^{+0.01}_{-0.01}$ & 0.95 \\[0.20cm]
\hline

2022 May 22 & 59721.6097 & \textit{NICER} & 2518 & 5202230217 & \phantom{0}$0.242^{+0.00309}_{-0.00337}$ & - & \phantom{0}$1.89^{+0.02}_{-0.02}$ & 0.99 \\[0.20cm]
\hline

2022 May 24 & 59723.4197 & \textit{NICER} & 1698 & 5202230218 & \phantom{0}$0.148^{+0.00412}_{-0.00303}$ & - & \phantom{0}$1.99^{+0.03}_{-0.03}$ & 1.13 \\[0.20cm]
\hline

2022 May 28 & 59727.5261 & \textit{Swift/XRT} & 850 & 00014374013 & \phantom{0}$0.271^{+0.0304}_{-0.0288}$ & - & \phantom{0}$1.78^{+0.12}_{-0.12}$ & 0.74 \\[0.20cm]

& 59727.8056 & \textit{NICER} & 2467 & 5202230219 & \phantom{0}$0.309^{+0.00376}_{-0.00373}$ & - & \phantom{0}$2.01^{+0.02}_{-0.02}$ & 0.94 \\[0.20cm]
\hline

2022 May 31 & 59730.3875 & \textit{NICER} & 1086 & 5202230220 & \phantom{0}$0.557^{+0.00642}_{-0.0112}$ & - & \phantom{0}$1.97^{+0.02}_{-0.02}$ & 1.06 \\[0.20cm]
\hline

2022 Jun 05 & 59735.6037 & \textit{NICER} & 1392 & 5202230221 & \phantom{0}$1.660^{+0.0133}_{-0.0140}$ & - & \phantom{0}$1.87^{+0.01}_{-0.01}$ & 1.41 \\[0.20cm]
\hline

2022 Jun 06 & 59736.9379 & \textit{Swift/XRT} & 787 & 00014374014 & \phantom{0}$1.829^{+0.211}_{-0.203}$ & - & \phantom{0}$1.86^{+0.13}_{-0.13}$ & 1.11 \\[0.20cm]
\hline

2022 Jun 08 & 59738.3775 & \textit{NICER} & 2178 & 5202230222 & \phantom{0}$3.980^{+0.0127}_{-0.0181}$ & \phantom{0}$0.171^{+0.00486}_{-0.00478}$ & \phantom{0}$1.79^{+0.01}_{-0.01}$ & 1.10 \\[0.20cm]
\hline

2022 Jun 11 & 59741.9898 & \textit{NICER} & 1642 & 5202230223 & \phantom{0}$7.350^{+0.0276}_{-0.0309}$ & \phantom{0}$0.209^{+0.00397}_{-0.00393}$ & \phantom{0}$1.83^{+0.01}_{-0.01}$ & 1.19 \\[0.20cm]
\hline

2022 Jun 14 & 59744.3546 & \textit{Swift/XRT} & 697 & 00014374015 & \phantom{0}$5.620^{+0.102}_{-0.102}$ & - & \phantom{0}$2.07^{+0.03}_{-0.03}$ & 1.34 \\[0.20cm]
\hline

2022 Jun 19 & 59749.0817 & \textit{Swift/XRT} & 102 & 00014374016 & \phantom{0}$1.620^{+0.589}_{-0.473}$ & - & \phantom{0}$1.69^{+0.37}_{-0.36}$ & 1.01 \\[0.20cm]

& 59749.4736 & \textit{NICER} & 1967 & 5202230224 & \phantom{0}$1.590^{+0.0119}_{-0.00920}$ & - & \phantom{0}$1.84^{+0.01}_{-0.01}$ & 1.42 \\[0.20cm]
\hline

2022 Jun 21 & 59751.9903 & \textit{NICER} & 673 & 5202230225 & \phantom{0}$0.617^{+0.0114}_{-0.0115}$ & - & \phantom{0}$2.01^{+0.02}_{-0.02}$ & 1.26 \\[0.20cm]
\hline

2022 Jun 22 & 59752.2484 & \textit{NICER} & 1082 & 5202230226 & \phantom{0}$0.705^{+0.00932}_{-0.0103}$ & - & \phantom{0}$1.87^{+0.02}_{-0.02}$ & 1.08 \\[0.20cm]
\hline

2022 Jun 25 & 59755.5594 & \textit{Swift/XRT} & 972 & 00014374017 & \phantom{0}$0.607^{+0.0616}_{-0.0592}$ & - & \phantom{0}$1.66^{+0.12}_{-0.12}$ & 1.08 \\[0.20cm]
\hline

2022 Jun 29 & 59759.2156 & \textit{NICER} & 789 & 5202230227 & \phantom{0}$0.365^{+0.00688}_{-0.00952}$ & - & \phantom{0}$2.04^{+0.03}_{-0.03}$ & 0.96 \\[0.20cm]
\hline

2022 Jul 03 & 59763.7992 & \textit{NICER} & 180 & 5202230228 & \phantom{0}$0.250^{+0.0177}_{-0.0160}$ & - & \phantom{0}$2.02^{+0.09}_{-0.08}$ & 1.20 \\[0.20cm]
\hline

2022 Jul 04 & 59764.3262 & \textit{NICER} & 692 & 5202230229 & \phantom{0}$0.405^{+0.0105}_{-0.00843}$ & - & \phantom{0}$1.56^{+0.03}_{-0.03}$ & 1.33 \\[0.20cm]
\hline

2022 Jul 05 & 59765.5123 & \textit{Swift/XRT} & 907 & 00014374018 & \phantom{0}$0.237^{+0.0202}_{-0.0199}$ & - & \phantom{0}$1.88^{+0.10}_{-0.10}$ & 1.36 \\[0.20cm]
\hline

2022 Jul 09 & 59769.7608 & \textit{Swift/XRT} & 693 & 00014374019 & \phantom{0}$0.0436^{+0.0117}_{-0.0101}$ & - & \phantom{0}$2.20^{+0.35}_{-0.32}$ & 0.84 \\[0.20cm]
\hline

2022 Jul 13 & 59773.1490 & \textit{NICER} & 1007 & 5202230230 & \phantom{0}$0.013^{+0.00294}_{-0.00150}$ & - & \phantom{0}$2.67^{+0.02}_{-0.02}$ & 1.83 \\[0.20cm]
\hline

2022 Jul 20 & 59780.2470 & \textit{NICER} & 1636 & 5202230232 & \phantom{0}$0.011^{+0.00313}_{-0.00276}$ & - & \phantom{0}$1.52^{+0.04}_{-0.03}$ & 1.01 \\[0.20cm]

& 59780.6309 & \textit{Swift/XRT} & 1052 & 00014374021 & \phantom{0}$0.00263^{+0.00440}_{-0.00162}$ & - & \phantom{0}$3.05^{+1.29}_{-1.27}$ & 0.78 \\[0.20cm]
\hline

\end{longtable}

\end{center}

\bsp	
\label{lastpage}
\end{document}